\documentclass[twocolumn,aps,floats,superscriptaddress,showpacs]{revtex4}
\usepackage{amsmath}

\usepackage{epsfig}
\usepackage{graphicx}
\usepackage{dcolumn}
\usepackage{bm}

\def\gapp{\lower.35em\hbox{$\stackrel{\textstyle>}{\sim}$}}
\def\lapp{\lower.35em\hbox{$\stackrel{\textstyle<}{\sim}$}}

\begin{document}
\bibliographystyle{apsrev}
%


\title{Dislocations and torsion in graphene and related systems }

\author{Fernando de Juan }
\affiliation{Instituto de Ciencia de Materiales de Madrid,\\
CSIC, Cantoblanco; 28049 Madrid, Spain.}
\author{Alberto Cortijo}
\affiliation{Department of Physics, Lancaster University,
Lancaster, LA1 4YB, United Kingdom.}
\author{Mar\'{\i}a A. H. Vozmediano}
\affiliation{Instituto de Ciencia de Materiales de Madrid,\\
CSIC, Cantoblanco; 28049 Madrid, Spain.}

\date{\today}
\begin{abstract}
A continuum model to study the influence of dislocations on the
electronic properties of condensed matter systems is described and
analyzed. The model is based on a geometrical formalism that
associates a density of dislocations with the torsion tensor and
uses the technique of quantum field theory in curved space. When
applied to two-dimensional systems with Dirac points like graphene
we find that dislocations couple in the form of vector gauge
fields similar to these arising from curvature or elastic strain.
We also describe the ways to couple dislocations to normal metals
with a Fermi surface.

\end{abstract}
%
%
%
%
 \maketitle

\section{Introduction}
Since the experimental synthesis of graphene the interest in this
material has mostly focussed on its electronic properties
\cite{NGetal09}. As time goes by and as the experiments become
more accurate the interest is being displaced to the morphological
aspects and their possible influence on the electronic properties
\cite{GBetal08}. Part of the interest on these issues is prompted
by the necessity to find the mechanisms limiting the mobility of
the graphene samples \cite{KG08}, an important issue for the
potential technological applications. Another source of interest
lies on the special mechanical and elastic properties of graphene
recently tested in experiments
\cite{LWetal09,DBetal09,BMetal09,TLetal09,SJetal09} and on the
fascinating possibility to explore the physics of real two
dimensional crystals.

Local disorder in graphene has been thoroughly studied  and we
refer to the review article \cite{NGetal09} for a fairly complete
list of references. A different type of disorder is provided by
the observation of ripples  in suspended graphene
\cite{Metal07,Metal07b} and in graphene grown on a substrate
\cite{Setal07,Ietal07}. The elastic and mechanical properties of
graphitic structures have been studied intensely in the past,
mostly in the context of understanding the formation of fullerenes
and nanotubes.  Very little work has been done for the flat
graphene sheet \cite{FLK07,CK07} and topological defects have been
often excluded in these studies. In the fullerene literature it
was established that the formation of topological defects
(substitution of a hexagonal ring by other polygons) is the
natural way in which the graphitic net heals vacancies and other
damages produced for instance by irradiation \cite{LWetal05}.
Among those, disclinations (isolated pentagon or heptagon rings),
dislocations (pentagon-heptagon pairs) and Stone-Wales (SW)
defects (special dislocation dipoles) were found to have the least
formation energy and activation barriers. Dislocations and SW
defects have been observed in carbon structures \cite{Hetal04} and
are known to have a strong influence on the electronic properties
of nanotubes. Dislocations in irradiated graphitic structures have
been described in \cite{EHB02} and experimental observations have
been reported in graphene grown on Ir in \cite{CNetal08}.

In previous works we used a general relativity formalism to  study
the influence of curved portions of the lattice on the electronic
properties of graphene by coupling the Dirac fermions to the
corresponding curved space. We found that curvature generates a
fictitious gauge field - a result that is also obtained in  more
conventional descriptions - and, in addition, we predicted a
space-dependent parameter in the kinetic term (similar to a
space-dependent Fermi velocity) that depends on the intrinsic
curvature of the sample and has non-trivial effects on the density
of states  \cite{CV07a,CV07b,JCV07,VJC08} and on the conductivity
\cite{CV09}. In the present work we extend this formalism to
include a finite density of  dislocations. Dislocations are
modelled by adding torsion to the graphene sheet either curved or
flat.

Torsion was first considered by Cartan and then by Sciama and
Kibble in order to deal with spin in General Relativity
\cite{Hetal76}. Within this scheme the spin of a particle turns
out to be related to the torsion just as its mass is related to
the curvature. Torsion is generally ignored in the context of GR,
because the torsion field is algebraic in the Einstein-Cartan
theory (its equations of motion are constraints, it has no
dynamics) and as a consequence there is no torsion in the absence
of matter. When matter generating torsion is added, its observable
effects are suppressed by the smallness of the gravitational
coupling and they are negligible \cite{S02}. These two drawbacks
are overcome when torsion is included in a geometric fashion as
associated to dislocations. In the context of electron systems
like graphene the dynamics of the lattice defects occurs at a much
higher energy than the electronic processes (in graphene, lattice
processes are related to the sigma bonds and have typical energies
of the order of tens of eV while the continuum relevant low energy
processes are of few tens of meV). It then makes perfect sense to
consider the motion of electrons in a frozen geometry and hence
the fact that torsion has no dynamics is not a problem. Moreover
the suppression of the coupling of the torsion and electronics
degrees of freedom will be of the order of magnitude of the
quotient between electronic and elastic degrees of freedom that,
although small,  will be much bigger than the gravitational scale.

The purpose of these notes is twofold: First we describe a
continuum description that allows to explore the effects of
dislocations on the electronic properties of graphene and other
materials, and, second, we propose an ``analog" model that
amplifies the effect of torsion in a General Relativity context.

\section{Dislocations and torsion}
\label{sectdisloc} 
\begin{figure}
\begin{center}
\includegraphics[width=4.5cm]{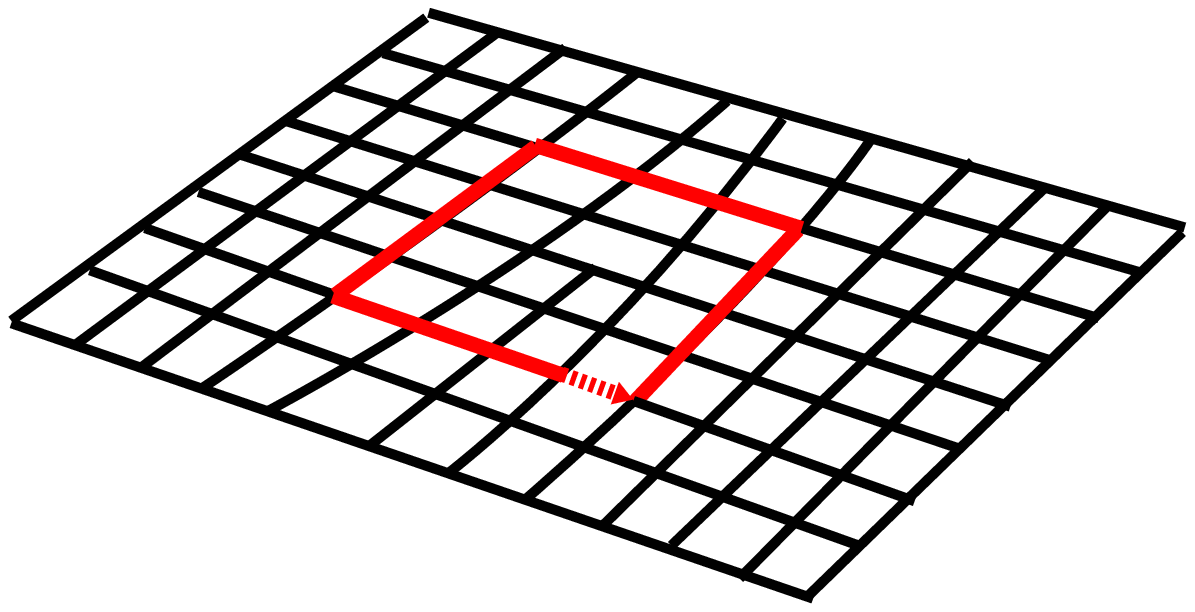}
\hspace{1cm}
\includegraphics[width=4.5cm]{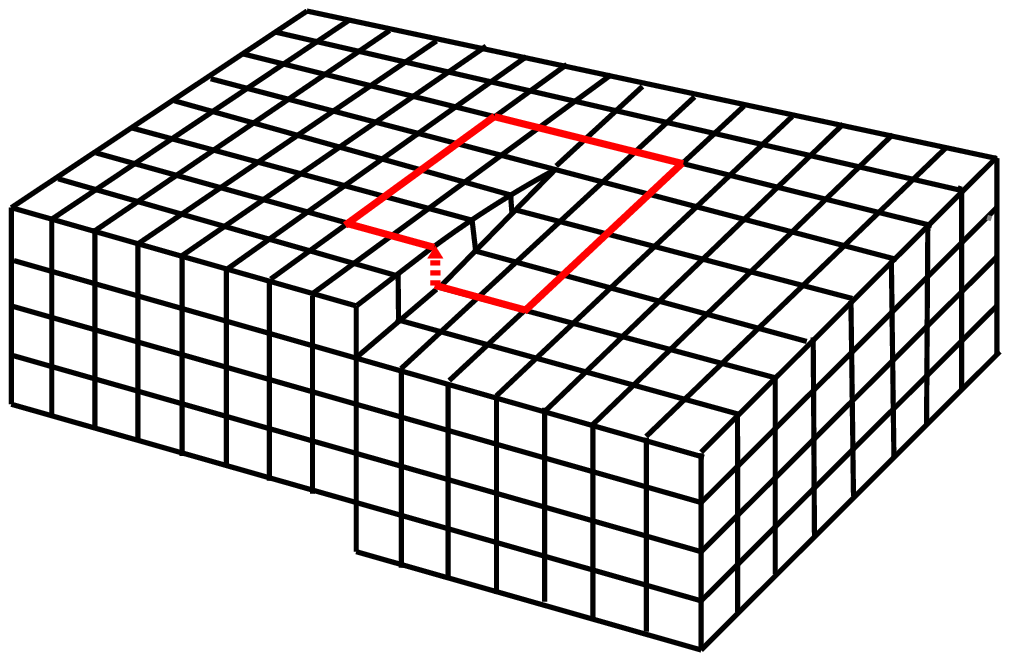}
\caption{(Color online) Upper: Edge dislocation in a square
two-dimensional lattice. The Burgers vector is parallel to the
displacement. Down: Screw dislocation in a three dimensional cubic
lattice. The Burgers vector is perpendicular to the displacement.}
    \label{lattice}
\end{center}
\label{disloc}
\end{figure}

\begin{figure}[h]
\begin{center}
\includegraphics[width=4.5cm]{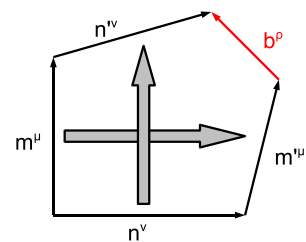}\label{burgers}
\caption{Non-closure of infinitesimal parallelograms due to the
presence of torsion}
\end{center}
\end{figure}

Topological defects in crystal lattices can be
classified in two kinds: Dislocations and disclinations. A
dislocation can be thought of being formed by performing a cut in
a bulk material and introducing extra rows of atoms (Volterra
process). These types of point defects can be  easily identified
by going around a closed circuit enclosing the end of the line cut
with discrete lattice steps. Without defects the path will end up
at the same point. In the presence of a dislocation it ends up
somewhere else and an extra vector has to be added to close the
path. This is called the Burgers vector. Dislocations can be
further classified by the direction of the Burgers vector with
respect to the plane defined by the circuit: In an edge
dislocation the Burgers vector lies in this plane, while in screw
dislocation it is perpendicular to it.  Fig. \ref{lattice} shows
an edge a and screw dislocation in the square lattice
characterized by a Burgers vector  parallel (perpendicular) to the
displacement. A disclination is similarly formed by adding a wedge
of atoms, and it can be identified by computing the total angle
subtended by a closed path enclosing the defect (Frank angle). A
disclination dipole can also be seen as a dislocation.
>From the point of view of geometry, dislocations represent a
source of translational mismatch ``anholonomy",(see Appendix
\ref{geometry} for a rigorous description of torsion), and as we
will see, are very naturally described in the continuum by a
connection with torsion.
In differential geometry, a connection defines the notion of
parallel transport. When a vector $B^\mu$ is parallel transported
from $x^\mu$ to $x^\mu+dx^\mu$, it experiences an infinitesimal
variation given by
\begin{equation}
\delta B^\mu=-\Gamma_{\nu\rho}^\mu (x)B^\nu dx^\rho.
\end{equation}
The equivalent of the closed circuit in the definition of the
Burgers vector can be defined by taking two infinitesimal vectors
$m^{\mu}$ and $n^{\nu}$, and trying to build a parallelogram with
them. For this purpose, one parallel transports the vector
$m^{\mu}$ along $n^{\nu}$:
\begin{equation}
m'^{\rho}=m^{\rho}+n^{\rho}-\Gamma^{\rho}_{\mu\nu}m^{\mu}n^{\nu},
\end{equation}
and the vector $n^{\nu}$ along $m^{\mu}$
\begin{equation}
n'^{\rho}=m^{\rho}+n^{\rho}-\Gamma^{\rho}_{\nu\mu}m^{\mu}n^{\nu}.
\end{equation}
As depicted in Fig. \ref{burgers}, if there is a dislocation in
the region enclosed by the path, the parallelogram obtained does
not close, and the part that is missing is proportional to the
antisymmetric part of the connection, defined as the torsion:
\begin{equation}
m'^{\rho}-n'^{\rho} = \left[
\Gamma^{\rho}_{\mu\nu}-\Gamma^{\rho}_{\nu\mu}\right]
m^{\mu}n^{\nu} \equiv T^{\rho}_{\mu\nu}m^{\mu}n^{\nu}
\end{equation}
Notice that the two  paths would have closed in a curved space
with a standard affine connection.

In the continuum limit, one can describe a density of dislocations
by the density of Burgers vector, and we see that torsion plays
precisely this role. Hence the natural identification of torsion
with dislocations.
The connection of torsion with the continuum theory of crystal
dislocations goes back to Kondo \cite{K52} and has been formalized
in \cite{KV92,K89}. Various aspects of the problem have been
explored in \cite{FM99}. A nice review on the relation of gravity
with topological defects in solids is \cite{M00}. The geometric
approach to defects in solids \cite{K89,SN88} relates the metric
of the curved -crystalline- surface with the deformation tensor
and establishes that disclinations are associated to the curvature
tensor and a finite density of dislocations generates a torsion
term.

\begin{figure}
\begin{center}
\includegraphics[width=4.5cm]{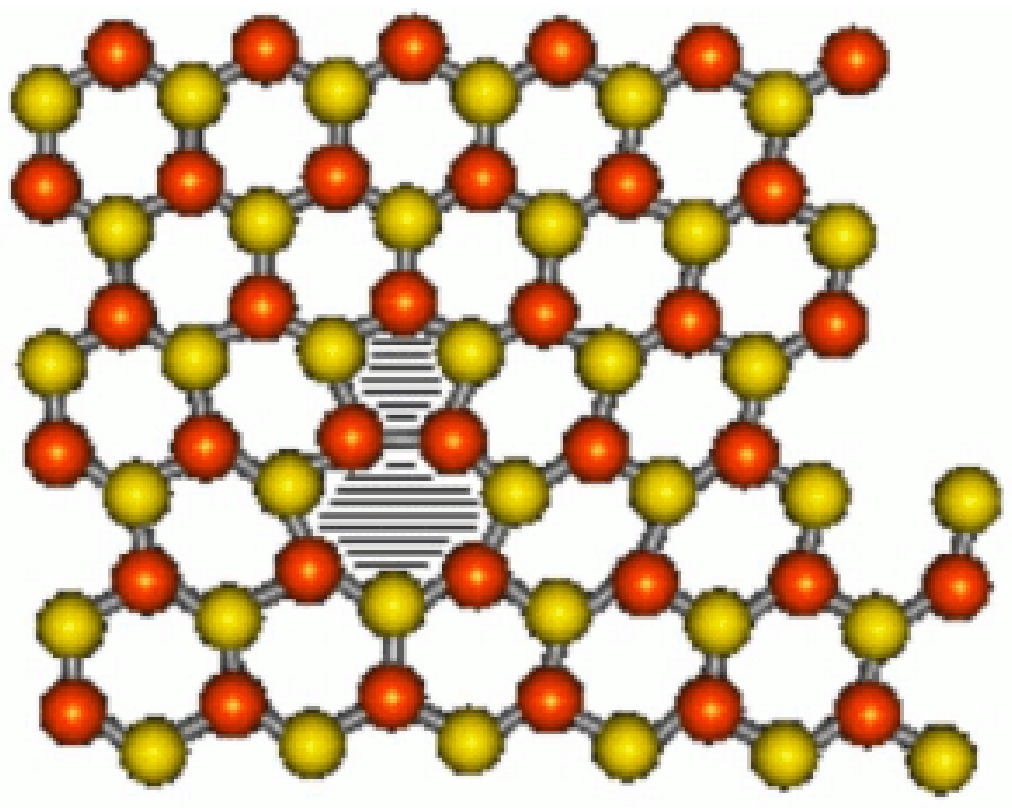}
\hspace{1cm}
\includegraphics[width=4.5cm]{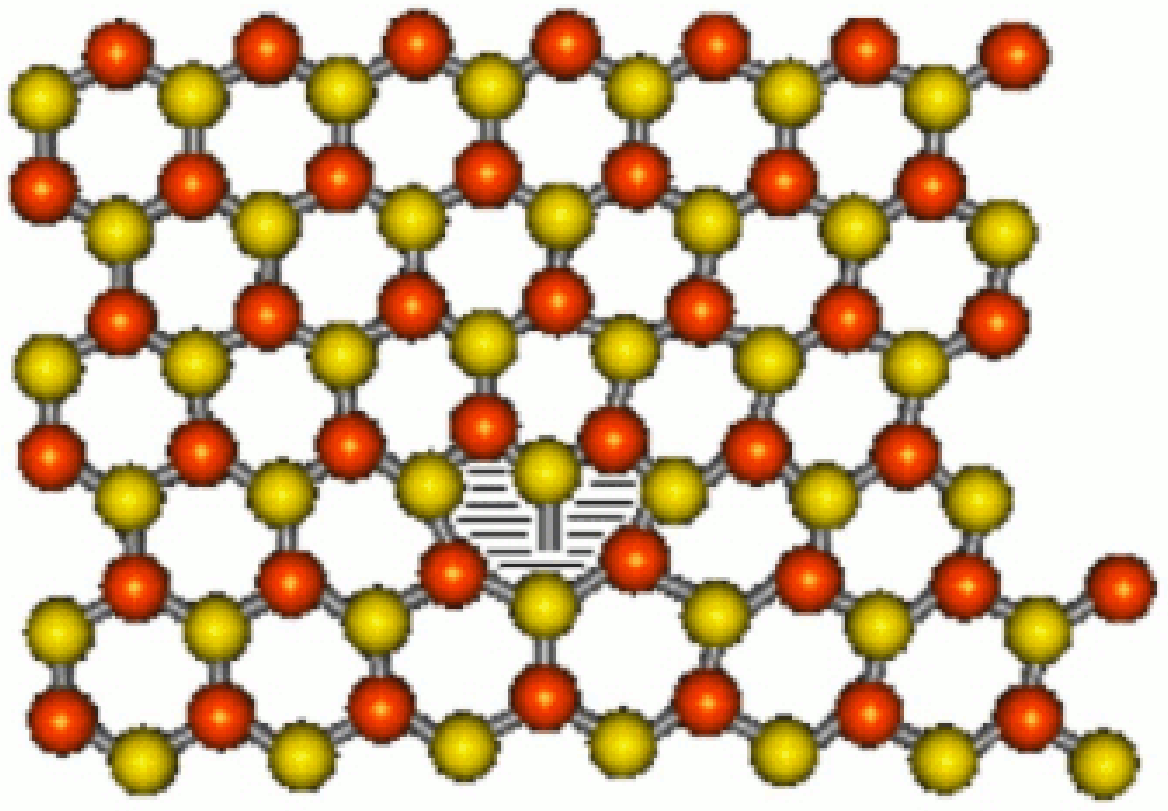}
\caption{(Color online) Structure of the glide (up) and shuffle
(down) dislocations in the planar graphene lattice.}
\end{center}
\label{dislocgraph}
\end{figure}

In flat surfaces screw dislocations do not exist. Dislocations in
graphene are made of pentagon--heptagon pairs and they have been
widely studied in connection with the properties of carbon
nanotubes \cite{SDD98} and, more recently, in the flat graphene
sheets or ribbons \cite{Aetal94,Cetal08,LJV09}. Observations of
topological defects in graphene have been reported in
\cite{Aetal01,Hetal04,DML04}. The stability of dislocations in
graphene has been explored recently in \cite{Cetal08} and two
types of stable dislocations were found. The most common are the
so-called glide dislocations made by a pentagon-heptagon pair
shown in the upper part of Fig. \ref{dislocgraph}. In addition we
can have shuffle dislocations (lower part of Fig.
\ref{dislocgraph} made by a octagon with a dangling bond. Both are
edge dislocations and add one or several lines of atoms to the
lattice. The difference is that shuffle dislocations contain an
extra atom with respect to the glide dislocations. In an infrared
continuum formulation both are seen as point defects that do not
disturb the lattice out of the core. The difference with conical
defects as pentagons or heptagons is that when an electron circles
such a defect it will found itself displaced from the original
position a distance proportional to the Burgers vector as shown in
Fig. \ref{disloc}.
\section{Coupling Dirac fermions to a curved space with dislocations}

The model that we propose to study the low energy properties of
graphite and graphene in the presence of lattice defects
(disclinations and dislocations) is based on the assumption that
the  low--energy   description of the quasiparticles in terms of
massless Dirac fermions remains valid in the presence of a diluted
distribution of defects. The stability of the Fermi points in
multilayer arrangements of the honeycomb lattice has been
demonstrated in \cite{MGV07}. As we have described in the previous
section the presence of dislocations in a crystal can be described
in the continuum limit by adding torsion to the space spanned by
the lattice. We will here follow a covariant approach close to the
general relativity formulation what allows to consistently couple
matter to geometry. This description began with the early paper of
ref. \cite{GGV92} where the Dirac equation on a sphere was used to
solve the electronic spectrum of fullerenes. The pentagonal
defects (disclinations) were described by a fictitious gauge
field. The formalism was later generalized to study the influence
of disclinations \cite{CV07d,CV07e} and of smooth curvature
\cite{JCV07,VJC08} on the density of states of graphene. These
previous works showed that the covariant formulation allows to
obtain sensible predictions that are in agreement with the
physical observations when the infrared properties are considered.
In particular it was shown that the curvature  induces a
fictitious vector gauge field. We will here analyze the nature of
the effective gauge fields created by the presence of dislocations
within the covariant framework.

Before introducing the changes induced by having torsion we
summarize the situation in a curved space.

The dynamics of a massless Dirac spinor in a curved spacetime is
governed by the modified Dirac equation (see Appendix
\ref{diractorsion}):
\begin{equation}
i\gamma^{\mu}({\bf r})D_{\mu}\psi=0 \label{dircurv}
\end{equation}
where $\gamma^{\mu}$ are  the curved space $\gamma$ matrices that
depend on the point of the space and can be computed from the
generalized anticommutation relations
\begin{equation}
\{\gamma^{\mu}({\bf r}),\gamma^{\nu}({\bf r})\}=2g^{\mu\nu}({\bf
r}),\nonumber
\end{equation}
and the covariant derivative operator is defined as
$$
D_\mu=\partial_{\mu}-\Omega_{\mu},
$$
where $\Omega_{\mu}$ is the spin connection of the spinor field
whose physical implications have been discussed in
\cite{JCV07,VJC08}.

Coupling a Dirac field to a curved space with torsion gives the
so-called  Einstein-Cartan-Dirac theory. The details of the
derivation of the Dirac equation in a space with torsion are given
in Appendix \ref{torsion}. Here we will simplify the description
putting it in a different, more physical way. The minimal coupling
of any geometrical or physical fields to the Dirac spinors adopts
always the form of a covariant derivative:
\begin{equation}
\partial_\mu \Rightarrow D_\mu=\partial_\mu + A_\mu ,
\end{equation}
where the given vector can be an electromagnetic potential induced
by a real electromagnetic field or any other real or fictitious
gauge field associated to deformations or to geometrical factors
\cite{VKG09}. In the case of having a density of dislocations in
the graphene sheet modelled by torsion we can construct  two
potential candidates to gauge fields. In four dimensions, from the
rank three torsion tensor (\ref{torsiontensor}) $T_{\mu\nu\rho}$
the following vectors can be built:\begin{equation}
V_\mu=g^{\nu\rho}T_{\nu\rho\mu}\;\;,\;\;
S_\mu=\epsilon_{\mu\nu\rho\sigma}T_{\nu\rho\sigma}.
\label{vectors}
\end{equation}
The field $V_\mu$ is a real vector and can be associated to the
density of edge dislocations while $S_\mu$ is an axial (pseudo)
vector associated to the density of screw dislocations. These
fields couple to the vector and axial current density respectively
so the full  Lagrangian is:
\begin{equation}
L_{{\rm int}}=\int d^4x \bar\Psi\;[\gamma^\mu (\partial_\mu +i e
V_\mu+ i\eta \gamma^5 S_\mu)]\Psi, \label{torsionint}
\end{equation}
where $e$ and $\eta$ are coupling constants related to the density
of edge and screw dislocations respectively.  The physical
consequences of the vector torsion are then similar to these
coming from curvature or elasticity. The coupling of the spinor to
the axial field $A_\mu$  can have important consequences as it
breaks time reversal symmetry.

\subsection{Two dimensional systems}
In (2+1) dimensions the completely antisymmetric part of the
torsion tensor is proportional to the Levi-Civita tensor:
\begin{equation}
T_{\mu\nu\rho}=\varepsilon_{\mu\nu\rho} \Phi, \label{2scalar}
\end{equation}
where $\Phi$ is a pseudoscalar field.

In 2+1 dimensions there are no screw dislocations, simply because
there is no third spatial dimension, and the antisymmetric part of
the torsion is rather related to time displacements. Since the
time components of the torsion generated by a dislocation are
zero, this term vanishes always. The remaining coupling to the
trace part acts a new vector gauge field with special symmetry
properties associated to the dislocations structure of the sample.
This field will add up to these coming from curvature or to real
electromagnetic fields.

A remaining interesting question is that of how this extra gauge
field couples to the electronic excitations around different Fermi
points (valleys). It is known that the fictitious gauge fields
arising from elastic deformations as well as these coming from
curvature, give rise to effective magnetic fields pointing in
opposite directions in different valleys. The reason is that the
sign of the coupling of the effective vector fields depends
uniquely on the definition of the spin connection (\ref{scon})
\begin{equation}
\Omega_{\mu}=\frac{1}{8}\omega_{\mu}^{\;
ab}\left[\gamma_a,\gamma_b\right], \label{scon}
\end{equation}
and is determined by the product of gamma matrices in it. As the
effective Hamiltonians around each of the Fermi points differ in
the sign of only one of the gamma matrices (they are similar to
parity conjugated) the spin connection has opposite signs in the
two valleys. This is irrespective of whether the connection has or
not an antisymmetric part and hence applies also to the coupling
induced by torsion.

\subsection{Three dimensional systems supporting Dirac fermions}
\label{3D} We have seen in the previous sections that the effects
of dislocations in two dimensional crystals like graphene are --
at most -- describable in terms of a fictitious gauge field
similar to the one associated to curvature. The formalism outlined
in this work acquires full power when applied to three dimensional
crystals supporting screw dislocations and having Dirac fermions
in the spectrum. Graphite is one of the most obvious candidates
where screw dislocations are very common \cite{H65} and elementary
excitations behaving as Dirac fermions were predicted near the K
points of the band structure in the early calculations
\cite{W47,SW58}. These Dirac fermions have been experimentally
confirmed in \cite{ZGetal06,Oetal08}.

The other perfect candidates are the the topological insulators
\cite{JM09,XQetal09} specially the $Z_2$ type that are protected
by time reversal symmetry. The axial part of the torsion will
couple to the three dimensional Dirac fermions, break the TRS and
affect the topological stability. The density of states associated
to the screw dislocations in \cite{ZRV09} can be reproduced with
the techniques outlined in this work and will be reported
elsewhere \footnote{F. de Juan, A. Cortijo, and M. A. H.
Vozmediano, work in progress.}.

\section{Dislocations in conventional electronic systems} Another
interesting question is how dislocations affect the electronic
properties of crystal systems whose quasiparticles are described
by the conventional Schrodinger equation. Within the present
formalism the question can be posed as: Do scalar fields see
torsion?

The question is interesting and relevant since the majority of
electronic systems (Fermi liquids) obey a conventional Schrodinger
equation and can be modelled with scalar fields. The point  to
notice here is that, although scalar fields transform in quite a
trivial manner under rotations and translations, building a proper
Hamiltonian means building derivatives and there is where
curvature and perhaps torsion plays a role. The Lagrangian for a
scalar field in a curved space is
\begin{equation}
\mathcal{L}_0=\int d^{n}x
\frac{1}{2}\sqrt{g}g^{\mu\nu}\partial_{\mu}\phi\partial_{\nu}\phi.
\end{equation}
As we see it involves only the metric and hence it does not see
the torsion. In the spirit of building phenomenological couplings
of the electronic density  to the dislocations the most natural
scalar couplings that we can form from the torsion vectors
(\ref{vectors}) are
\begin{equation}
\mathcal{L}_{{\rm int}}=\int d^{n}x
g_1(\partial_{\mu}V^{\mu})\phi^2+g_2 V^{\mu}V_{\mu}\phi^2,
\end{equation}
where $V_\mu$ is the vector associated to the trace of the
torsion, i. e., to the density of edge dislocations. The coupling
$g_2$ being quadratic in the torsion will be suppressed with
respect to $g_1$ by an extra power of the density of dislocations.
In the two dimensional case we can form the same interactions with
the scalar field of (\ref{2scalar}).

Landau levels in the presence of a screw dislocation have been
explored with a similar formalism in \cite{FM99}.

\section{Conclusions and discussion}
The proposal of modelling  dislocations in a crystal with a
geometrical approach including torsion within the elasticity
theory is very old. What is new in our approach is the suggestion
to study the influence of dislocations on the electronic
properties of graphene and other electronic systems having  Dirac
fermions in the spectrum, by coupling the Dirac equation to a
generalized curved space with torsion. In what two dimensional
graphene sheets are concerned the main results of this work are
somehow disappointing: spinors in two spatial dimensions do not
couple to the axial part of the torsion (since there are no screw
dislocations). Such a coupling would have had very important
consequences.  The coupling to the trace part is in the form of a
vector gauge field similar to the one produced by the curvature or
by strain in a more conventional approach. The spacial
distribution of this new gauge field will nevertheless depend on
the density of dislocations.

The most interesting applications of the formalism outlined here
will be in the area of topological insulators that often support
massless  chiral fermions in three dimensions. In \cite{AS99} it
was shown that the coupling of spinors to the axial vector induced
by torsion generates at higher order a four Fermi interaction on
the fermions what had an influence on neutrino oscillations. In
the context of condensed matter such a coupling would renormalize
the Hubbard interaction or generate it if it was absent. The
presence of the torsion axial vector in the Dirac equation will
have also implications on various fundamental aspects of the
physics of condensed matter system described by Dirac fermions. As
it breaks time reversal symmetry, it can prevent weak
localization.    As for the conductivity, the disorder induced by
a random distribution of dislocations may change the universality
class of previous models based on vacancies or other T-preserved
disorder. The breakdown of time reversal symmetry will also lift
the quantum protectorate associated to the Fermi points
\cite{MGV07} and allow the opening of a gap although the torsion
itself will not explain the observations reported on gaps in
graphene \cite{GFetal07,LLA09}. The physical implications of the
effective gauge fields induced by dislocations on graphene have
been discussed in \cite{GGV01,M07,GHD08,FGK08}.

\appendix

\section{The Dirac equation in a space with torsion and curvature}
\label{diractorsion}
\subsection{Dirac equation in a curved space without torsion}
\label{curvature} The behavior of  spinors in curved spaces is
more complicated than that of scalar or vector fields because
their Lorentz transformation rules do not generalize easily to
arbitrary coordinate systems \cite{W96}. Instead of the usual
metric $g_{\mu\nu}$ we must introduce at each point $X$ described
in arbitrary coordinates, a set of locally inertial coordinates
$\xi_X^a$ and the vielbein fields  $e_\mu^a(x)$, a set of
orthonormal vectors labelled by $a$ that fixes the transformation
between the local and the general coordinates:
\begin{equation}
e_\mu^a(X)\equiv\frac{\partial \xi_X^a(x)}{\partial
x^\mu}\vert_{x=X}.
\end{equation}
The curved space gamma matrices $\gamma^\mu(x)$ satisfying the
commutation relations
\begin{equation}
\{\gamma_\mu\gamma_\nu\}=2g_{\mu\nu},
\end{equation}
are related with the constant, flat space matrices $\gamma^a$ by
\begin{equation}
\gamma^\mu(x)=e^\mu_a\gamma^a.
\end{equation}
The spin connection $\Omega_\mu(x)$ is defined from the fielbein
by
\begin{equation}
\Omega_\mu(x)=\frac{1}{4}\gamma_a\gamma_b
e^a_\lambda(x)g^{\lambda\sigma}(x) \nabla_\mu e^b_\sigma(x),
\label{spincon}
\end{equation}
where $\nabla_\mu$ is the covariant derivative acting on the
$e_\mu$ vectors as
\begin{equation}
\nabla_\mu e^a_\sigma=\partial_\mu
e_\sigma^a-\Gamma_{\mu\sigma}^\lambda e_\lambda^a.
\end{equation}
The spin connection for 1/2 spinors can be written as
\begin{equation}
\Omega_{\mu}=\frac{1}{8}\omega_{\mu}^{\;
ab}\left[\gamma_a,\gamma_b\right], \label{scon}
\end{equation}
where $\omega_{\mu}^{\; ab}$ are the spin connection coefficients
\begin{equation}
\omega_{\mu}^{\; ab}=e^a_{\; \nu} \left(\partial _{\mu} +
\Gamma^{\nu}_{\mu\lambda} \right)e^{b\lambda}.\label{scoef}
\end{equation}
In the flat case $\Gamma_{\mu\sigma}^\lambda$ is the usual affine
connection (Christoffel symbols)  related to the metric tensor by
\begin{equation}
\Gamma_{\mu\sigma}^\lambda=\frac{1}{2}g^{\nu\lambda}\{\frac{\partial
g_{\sigma\nu}}{\partial x^\mu}+\frac{\partial g_{\mu\nu}}{\partial
x^\sigma}-\frac{\partial g_{\mu\sigma}}{\partial x^\nu}\}.
\label{christoffel}
\end{equation}
Finally, the determinant of the metric needed to define a scalar
density lagrangian is given by
\begin{equation}
\sqrt{-g}=[\det(g_{\mu\nu})]^{1/2}\;=\;\det[e_\mu^a(x)].
\label{jacobian}
\end{equation}
These formulas apply to a curved space without torsion. The Dirac
equation is then written as
\begin{equation}
i\gamma^\mu D_\mu\psi=0, \label{curveddirac}
\end{equation}
where $D_\mu=\partial_{\mu}+\Omega_{\mu}$ is the covariant
derivative acting on the spinors in the curved space  and
$\gamma^\mu=\gamma^a e_a^\mu$ the curved gamma matrices.

\subsection{Non-singular torsion in differential geometry}
\label{geometry}

In differential geometry we consider a curved manifold, described
by a set of coordinates $x^{\mu}$. To this manifold we associate a
metric $g_{\mu\nu}$ which defines the distance locally. A notion
of parallel transport is also needed to compare vectors in
different points, in order to construct meaningful derivatives.
The connection and the metric are independent geometric objects
that are usually related by the so-called metricity condition
ensuring that the relative angles and magnitudes of vectors are
preserved under parallel transport. The metricity condition reads:
\begin{equation}
D_{\lambda}g_{\mu\nu}=\partial_{\lambda}g_{\mu\nu}- \Gamma_{\
\lambda \mu}^{\rho}g_{\rho \nu} - \Gamma_{\ \lambda
\nu}^{\rho}g_{\rho \mu}=0, \label{metricity}
\end{equation}
this is, the covariant derivative of the  metric vanishes. The
metricity condition fixes the symmetric part of the connection and
leaves the antisymmetric part undetermined. The most general
connection, compatible with the Riemannian metric, is
\begin{equation}
\Gamma_{\ \mu \nu}^{\lambda}= \left\{
\begin{array}{c}
    \lambda \\
    \mu \ \nu
\end{array}
\right\} +K_{\ \mu \nu}^{\lambda},
\end{equation}
where:
$ \left\{
\begin{array}{c}
    \lambda \\
    \mu \ \nu
\end{array}
\right\} $ are the Christoffel symbols and $K^{\lambda}_{ \mu\nu}$
is called the contortion tensor related with the torsion tensor
 by
\begin{equation}
K^{\lambda}_{\mu\nu}=\frac{1}{2}g^{\lambda\beta}(T_{\beta\mu\nu}+
T_{\mu\beta\nu}+T _{\nu\beta\mu}).
\end{equation}
The torsion $T^{\lambda}_{\mu\nu}$ is the antisymmetric part of
the connection:
\begin{equation}
T^{\lambda}_{\mu\nu}=\Gamma ^{\lambda}_{\mu\nu}-\Gamma
^{\lambda}_{\nu\mu}. \label{torsiontensor}
\end{equation}
This means that when choosing a connection for a metric space, we
are always free to choose the antisymmetric part of the connection
(the torsion) at will. Only the symmetric part is  fixed by the
metricity condition. The usual choice is to take the torsion equal
to zero, so that the metric alone determines all geometric
properties. This is, for example, the case of General Relativity.
The only restriction on K is that
\begin{equation}
K_{\lambda\mu\nu}=-K_{\nu\mu\lambda}
\end{equation}
hence in a d-dimensional space it has $\frac{1}{2}[d^2(d-1)]$
components. The differential geometry of a general curved space is
determined by the two tensor fields $g_{\mu\nu}$ and
$T_{\mu\nu}^\alpha$ (or, equivalently $K_{\mu\nu}^\alpha$).

In a space with torsion the shortest path (geodesics) and the
covariantly paralell path do not coincide and, as was mentioned in
Section \ref{sectdisloc}, parallelograms do not close.

\subsection{Torsion in a 2D flat space}

Aiming to a description of dislocations alone,  we study here the
special case of a space with torsion but no curvature, in 2+0
dimensions. This will be the spatial part of the space-time used
to describe graphene with dislocations as the Dirac equation in a
space with spatial torsion.

As we will see, life is very simple in this space. First of all,
since there is no curvature, there is a global coordinate system
in which the metric tensor is the identity. Or seen from the
tetrad perspective, there is a coordinate system in which the
tetrads are trivial, and thus the Lorentz connection and the
Einstein connection are equal. (Since the tetrads are trivial,
there is no difference between latin and greek indices now, and
moreover, since we are in 2+0, upper and lower indices are equal.)
In this particular case, with the tetrads chosen in this way, we
can say that the torsion in encoded in the connection.

Moreover, the antisymmetry of the torsion tensor in 2D immediately
implies that it can be written as:
\begin{equation}
T^{\lambda}_{\mu\nu}=\epsilon_{\mu \nu}b^{\lambda},
\end{equation}
where $b^{\lambda}$ is an arbitrary vector field,  basically the
density of Burgers vector. It could be thought that this vector
has some further constraints due to the Bianchi identities
\cite{K89}, but it can be checked that in 2D if
$R_{\mu\nu\rho\sigma}=0$ these identities are trivial.

\subsection{Dirac equation in a curved space with torsion}
\label{torsion}

As have seen  in appendix \ref{geometry}, inclusion of torsion in
the geometrical description simply changes the connection and
hence the covariant derivative.

It would be tempting then to propose that the Dirac equation in
the space with torsion and curvature is given by eq.
(\ref{dircurv}) with the appropriate connection but there are some
important subtleties that we will specify \cite{S02,H02}.

Consider the Dirac field in flat space. The massless, manifestly
hermitian Dirac Langrangian may be written as:
\begin{equation}\label{dirac1}
\mathcal{L}=\int d^{n}x
\frac{1}{2}\left[\bar{\psi}\gamma^{\mu}\partial_{\mu}\psi -
\partial_{\mu}\bar{\psi}\gamma^{\mu}\psi\right].
\end{equation}
Noting that:
\begin{equation}
\left(\partial_{\mu}\bar{\psi}\right)\gamma^{\mu}\psi =
\partial_{\mu}\left[\bar{\psi}\gamma^{\mu}\psi \right] -
\bar{\psi}\gamma^{\mu}\partial_{\mu}\psi,
\end{equation}
plus the fact that a total derivative doesn't affect the equations
of motion (with suitable boundary conditions), the Lagrangian
(\ref{dirac1}) can be traded for:
\begin{equation}\label{dirac2}
\mathcal{L}=\int d^{n}x \bar{\psi}\gamma^{\mu}\partial_{\mu}\psi.
\end{equation}
Indeed, the Dirac equation can be derived by variation  with
respect to $\bar{\psi}$ of any of those Lagrangians. The simpler
form (\ref{dirac2}) seems to be not hermitian,  but this is not a
problem since we can always introduce the boundary term back.

Now consider the same problem in a manifold with curvature  and
torsion. Consider the hermitian Lagrangian:
\begin{equation}
\label{dirac3} {\mathcal L}=\int d^{n}x \sqrt{g} \frac{1}{2}
\left[\bar{\psi}\gamma^{\mu}D_{\mu}\psi -
\left(D_{\mu}\bar{\psi}\right)\gamma^{\mu}\psi\right],
\end{equation}
with $D_{\mu}=\partial_{\mu}-\omega_{\mu bc}\gamma^{b}\gamma^{c}$.
Now the equivalent of the chain rule is:
\begin{equation}
\left(D_{\mu}\bar{\psi}\right)\gamma^{\mu}\psi =
D_{\mu}\left[\bar{\psi}\gamma^{\mu}\psi \right]
-\bar{\psi}D_{\mu}\left(\gamma^{\mu}\psi\right).
\end{equation}
Note the following important point: since the covariant
derivative contains gamma matrices, we can not just commute it to
make it act directly on the spinor field. The ``chain rule'' that
we have to use is really:
\begin{equation}
\left(D_{\mu}\bar{\psi}\right)\gamma^{\mu}\psi =
D_{\mu}\left[\bar{\psi}\gamma^{\mu}\psi \right]
-\bar{\psi}\gamma^{\mu}D_{\mu}\psi + 4\omega_{\mu bc}e^{\mu b}
\bar{\psi} \gamma^{c} \psi,
\end{equation}
where we have used:
\begin{equation}
\gamma^{b}\gamma^{c}\gamma^{a}=
\gamma^{a}\gamma^{b}\gamma^{c}+2\left(\gamma^{b}\eta^{ac}-\gamma^{c}\eta^{ab}\right).
\end{equation}

The commutation with the gamma matrices has introduced a new term,
the trace of the connection. This means that the equivalence of
the simpler form given in eq. (\ref{dirac2}) is not just obtained
by promoting the derivative to a covariant one, but it also
requires the introduction of this trace \cite{H02}.

In four dimensions, the following identity:
\begin{equation}\label{id1}
\gamma^{a}\gamma^{b}\gamma^{c}=\gamma^{a}g^{bc}+\gamma^{c}g^{ab}-
\gamma^{b}g^{ac}+i\epsilon^{abcd}\gamma^{5}\gamma^{d},
\end{equation}
and the definition of the connection (\ref{scon}) allows to
rewrite this as:
\begin{equation}
\mathcal{L}=\int d^{4}x \left(\bar{\psi}\gamma^{a}\partial_{a}\psi
+\frac{T_{abc}}{4}\bar{\psi}\gamma^{[a}\gamma^{b}\gamma^{c]}\psi\right),
\end{equation}
or:
\begin{equation}
\mathcal{L}=\int d^{4}x \left(\bar{\psi}\gamma^{a}\partial_{a}\psi
+i\epsilon^{abcd}T_{bcd}\bar{\psi}\gamma^{5}\gamma^{a}\psi\right),
\end{equation}
Which reveals the well known result that,  in General Relativity,
fermions only couple to the antisymmetric part of the torsion
\cite{H02}. In three dimensions using the  identity:
\begin{equation}
[\gamma^{a},\gamma^{b}]=-2\epsilon^{abc}\gamma_{c},
\end{equation}
we can write the action in the form
\begin{equation}
\mathcal{L}=\int d^{3}x \left(\bar{\psi}\gamma^{a}\partial_{a}\psi
+i\epsilon^{abc}T_{abc}\bar{\psi}\psi\right),
\end{equation}
from where the Dirac equation can be extracted directly to read
\begin{equation}
[\gamma^{a}\partial_{a} +i\epsilon^{abc}T_{abc}]\psi=0,
\label{direqtor}
\end{equation}

\vspace{1cm}

{\it Note added}: After completing this work we became aware of
ref. \cite{MSZ09} where similar issues are discussed. We think
that, although there is some obvious overlapping, the two works
entertain different points of view and can be seen as
complementary.

\vspace{0.5cm}

\acknowledgments MAHV thanks the hospitality of the  Centro de
Fisica do Porto where this work began and the  interesting
conversations held there with the Cosmology group, specially with
C. Herdeiro. AC acknowledges financial support from the European
Commision, Marie Curie Excellence Grant MEXT-CT-2005-023778. This
research was supported by the Spanish MECD grant
FIS2005-05478-C02-01 and FIS2008-00124.

\bibliography{Torsion}

\end{document}